\documentclass[conference]{IEEEtran}
\IEEEoverridecommandlockouts
\usepackage{cite}
\usepackage{amsmath,amssymb,amsfonts}
\usepackage{algorithmic}
\ifCLASSINFOpdf
   \usepackage[pdftex]{graphicx}
\else
   \usepackage[dvips]{graphicx}
\fi
\usepackage{textcomp}
\usepackage[dvipsnames]{xcolor}
\usepackage[caption=false,font=footnotesize]{subfig}
\usepackage{booktabs} %
\graphicspath{{pdf/}{jpeg/}{png/}{eps/}{figures/}}
\usepackage{stfloats}
\usepackage{balance}
\usepackage{soul}
\usepackage[normalem]{ulem}
\usepackage{cancel}
\usepackage{verbatim}
\usepackage{nohyperref}  %
\usepackage{url}  %
\usepackage{cases}
\usepackage{multirow}

\begin{document}
\bstctlcite{MyBSTcontrol}

\title{Scaling Blockchains to Support Electronic Health Records for Hospital Systems
}

\author{
    \IEEEauthorblockN{Alyssa Donawa}
    \IEEEauthorblockA{
        Department of Computer Science\\ 
        University of Kentucky\\
        Lexington, KY, USA\\
        adonawa@uky.edu
    }
    \and
    
    \IEEEauthorblockN{Inema Orukari}
    \IEEEauthorblockA{
        Department of Biomedical Engineering\\
        Washington University in St. Louis\\ 
        St. Louis, MO, USA\\
        inema.orukari@wustl.edu
    }
    
    \and

    \IEEEauthorblockN{Corey E. Baker}
    \IEEEauthorblockA{
        Department of Computer Science\\ 
        University of Kentucky\\
        Lexington, KY, USA\\
        baker@cs.uky.edu
    } 

}

\maketitle

\begin{abstract}
Electronic Health Records (EHRs) have improved many aspects of healthcare and 
allowed for easier patient management for medical providers. 
Blockchains have been proposed as a promising solution for supporting 
Electronic Health Records (EHRs), but have also been linked to scalability 
concerns about supporting real-world healthcare systems.
This paper quantifies the scalability issues and bottlenecks related to current blockchains and 
puts into perspective the limitations blockchains have with supporting healthcare systems. Particularly we show that well known blockchains such as Bitcoin, Ethereum, and IOTA cannot support transactions of a large scale hospital system such as the University of Kentucky HealthCare system and leave over 7.5M unsealed transactions per day. We then discuss how bottlenecks of blockchains can be relieved with sidechains, enabling well-known blockchains to support even larger hospital systems of over 30M transactions per day. We then introduce the Patient-Healthchain architecture to provide future direction on how scaling blockchains for EHR systems with sidechains can be achieved.
\end{abstract}


\section{Introduction} \label{section:intro}

The healthcare industry has come a long way from the days of keeping paper records and using fax machines as a primary source of communication. 
Now there are Electronic Health Records (EHRs), which are digital translations of a patient’s paper chart\cite{van2009inter, thakkar2006risks}. %
EHRs were created with the intention of not only containing the treatment information of patients, but to also facilitate the quality of the care process.
EHRs are capable of sharing information with other providers and organizations involved with a patient‘s care~\cite{EHRinfo}.

EHRs, although an improvement upon the method of 
keeping paper copies of health records, are still a headache for medical providers %
\cite{reisman2017ehrs}. 
Hospitals use different EHR software\footnote{B. Siwicki, ``Biggest EHR challenges for 2018: Security, interoperability, clinician burnout,'' Healthcare IT News, 2017,  \url{https://www.healthcareitnews.com/news/biggest-ehr-challenges-2018-security-interoperability-clinician-burnout}, Accessed: 10/3/19} which leads to a lack of \textit{interoperability}, ``the ability of different information systems to connect in a coordinated manner across organizational bounds''\cite{interoperabilityDef},
across all medical facilities \cite{reisman2017ehrs}. 
The lack of interoperability poses an issue in the case of a patient whose treatment spans across 
multiple geographic locations or medical facilities\cite{van2009inter}. 
When transferring between locations of treatment, the information that would be 
readily available at one location will not necessarily be available at a second 
location. 
Not having patient information promptly available could result in wasted resources due to the second location having a 
requirement to run tests on a patient that were already ran at the first 
facility. %
Another limitation with the current EHR process is %
patient access, %
as patients do not have immediate access to their complete EHR. 
In addition, providers do not need to ask for patient consent to share information for a patient's general care in the case of transfers. 
Not requiring consent means a patient will not always know who is accessing their health records~\cite{samhsa}. %
For patients, having quick access to a complete health record would not only keep them aware but also mitigate concerns about who is accessing and making changes to their EHRs.

EHRs have improved quality of care and made record management easier for 
medical providers \cite{ehrBenefits} and APIs have been developed for EHRs. 
For example, the Health Level 7 (HL7) standard Fast Healthcare Interoperability 
Resources (FHIR) is a RESTful open standard that has been adopted by many 
organizations.
FHIR provides a simple and consistent structure for health data across platforms~\cite{bender2013hl7,zhang2018fhirchain}.
Though FHIR greatly assists with the translation of data, there are still obstacles to 
overcome when it comes to privileges, access, accountability, and logging 
between hospital systems\cite{alhadhrami2017introducing}.

In the U.S., EHRs have to be in compliance with the \textit{Health Insurance Portability and Accountability Act} (HIPAA) regulations making the handling 
and transferring of health data a security risk. 
HIPAA regulations require that entities implement safeguards that protect the 
confidentiality, integrity, and availability of EHRs of 
patients~\cite{assistance2003summary}. 
Blockchains use cryptographic technologies to ensure confidentiality even on a public blockchain. 
Integrity of data is kept through chronological hashes on an immutable ledger, which also allows for the availability of information. %

From the perspective of a medical professional, EHRs allow for easier 
management of the large amount of patients that a medical provider deals with 
on a regular basis.
However, from a patient's perspective, there is still more work to be done to 
improve how they are able to manage and keep track of all of their health 
records.
Blockchains have been proposed~\cite{medrec, medshare, dubovit, bpds} %
 as a promising
solution to managing EHRs, but have been known to have scalability issues~\cite{linn2016blockchain,rifi2017towards,xia2017bbds}.
This paper will quantify the scalablity concerns with using blockchains for EHRs 
and discuss solutions to relieve the potential bottlenecks.
The contributions of this work are to: 
\begin{itemize}
    \item quantify and put into perspective the scalability issues related to current blockchains %
    \item discuss how to relieve bottlenecks using sidechains and highlight considerations
    \item propose the Patient-Healthchain architecture that uses sidechains and the Proof of Authority (PoA) consensus scheme to allow for high throughput and scalability %
\end{itemize}

The remainder of the paper is structured as follows:
Section~\ref{section:prelims} discusses relative preliminary information about aspects of blockchain technologies,
Section~\ref{section:related} discusses the related work,
Section~\ref{section:benefits} discusses the benefits of sidechains for EHRs, Section~\ref{section:architecture} describes the proposed architecture,  %
and Section~\ref{section:conclusion} concludes and mentions possible future work.

\section{Blockchain Preliminaries} \label{section:prelims} %
One of the most well known uses of blockchain technology is Bitcoin~\cite{nakamoto2008bitcoin}. 
Bitcoin is a cryptocurrency that attempts to decentralize the payment system. %
It removes the middle man of a financial institution and implements a peer-to-peer distribution of funds.
Not needing a middle man is achieved through the use of Proof of Work (PoW) %
and a ledger that stores a list of all transactions made via the blockchain.

A \textit{transaction} in a blockchain is an event that takes place and is 
then recorded on the blockchain.
All nodes in the network race to verify that these transactions are correct, 
creating a system of trust in an innately trustless environment.
A block is made up of a certain number of transactions before it is eventually 
mined to the blockchain. 
In the case of Bitcoin, a transaction is equivalent to a financial transaction.
\subsection{Notable Consensus Schemes}\label{subsection:consensus}

A consensus scheme is a way to validate a mined block before adding the block to the 
blockchain \cite{consensus}. 
PoW and Proof of Stake (PoS) are two well known consensus 
mechanisms used by Bitcoin and Ethereum respectively \cite{understandingConsensus, consensus}. %
PoW has been described as a ``cryptographic block-discovery racing game''\cite{consensus}.
In the case of Bitcoin, miners compete for bitcoin currency by solving hard hash puzzles that allow them to publish a block to the blockchain. The downside to PoW is that a lot of computing power is needed to solve the puzzles, which requires a huge electrical energy expenditure.

PoS was introduced to overcome the limitations of PoW \cite{understandingConsensus}. To achieve PoS, a miner's stake in the blockchain determines their mining power. Stake is determined by factors such as wealth or age in the blockchain, combined with random selection. %
A user who has been randomly selected as a validator will generate blocks at that given time. 
PoS reduces the energy and computing power needed to maintain PoW.
An alteration of PoS is Delegated Proof of Stake (DPoS), which is used by BitShares~\cite{dpos}. DPoS uses \textit{witnesses} and \textit{stakeholders} to validate and generate blocks. Stakeholders vote on witnesses to generate blocks in a round table fashion. If a witness does not perform as expected, the stakeholders can vote them out, allowing someone else to become a witness. This model allows for a more decentralized consensus process. DPoS does not require participation in block generating from all users in the network, which allows for less computing power than PoS.

PoA~\cite{poaNetwork} requires that a user who wants to be allowed to validate and generate blocks must request to do so. In order to get approved, a process is undertaken to verify identity. An example of this process would be getting a public notary to confirm an identity. Instead of solving complex math problems like in PoW, nodes in the network are designated as \textit{authorities} and only those authorities are allowed to generate and validate new blocks \cite{parityPoA}.%

Choosing the right consensus scheme is critical to producing a viable 
blockchain for use in healthcare. %
Traditional consensus schemes such as PoW or PoS may be suitable for a number 
of cases, but may not provide the best benefits for health systems.
Properties to determine the correct consensus scheme for a medical blockchain 
should consist of the following: low power consumption, throughput consistent 
with the average healthcare system, and scalability~\cite{blobel2006advanced}. 
\subsection{Smart Contracts} \label{subsection:smartcontracts}
A smart contract is a self running script that is coded onto the blockchain. %
Smart contracts automatically execute when the required conditions defined in the code are met.
Examples of these conditions are contractual terms being met, or a change in state~\cite{smartcontractApplications}.
A popular analogy used to describe smart contracts has been the vending machine comparison. 
When a person puts money into a vending machine, the person will either receive what they attempted to buy, or there will be an error and that person will get their money back. 

While popular use of smart contracts started with Ethereum~\cite{ethereum}, smart contracts have already been implemented across many decentralized applications in various ways.
Examples of smart contract applications include: defining trading criteria, access control, supply chain quality management, business process management, voting systems and identity management~\cite{smartcontractApplications}. 
In the context of medical applications, smart contracts can be used as an access control mechanism~\cite{contractAccessControl}. 

\subsection{Sidechains}\label{subsection:sidechains}
\begin{figure}%
    \includegraphics[width=\columnwidth, height=2.2in]{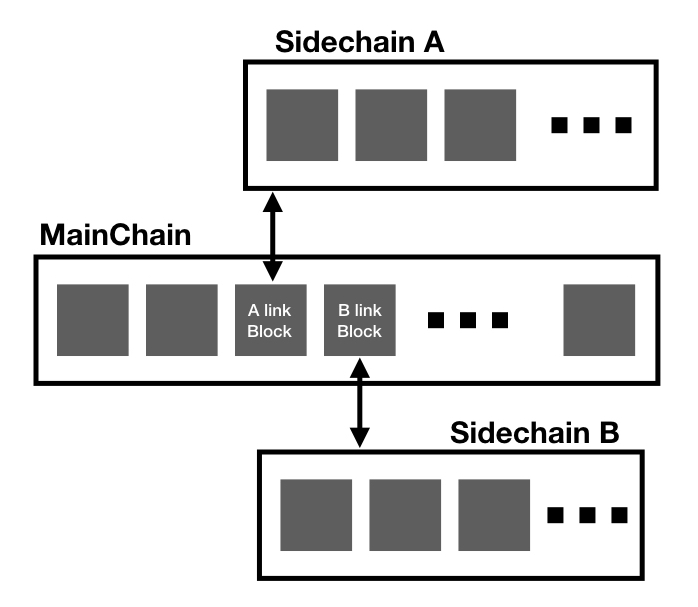}
  \caption{Mainchain and Sidechain Relationship}
  \label{fig:chain}
 \end{figure}

A sidechain is a blockchain separate from the parent blockchain or mainchain, 
connected via a two-way link~\cite{sidechains} demonstrated in Figure~\ref{fig:chain}. 
Sidechains offer the ability to offload tasks from the parent blockchain to the sidechain.
Sidechains have been used in Bitcoin; referred to as Altchains, and are used 
as a method to link and exchange other cryptocurrencies with bitcoins.
Sidechains also allow for testing of features before attempting to implement 
them to the mainchain.
One point to note about sidechains is that they operate as their own 
blockchain and therefore may require separate minors %
based on the selected consensus scheme.

Sidechains can also be used to complement a parent blockchain.
A newer blockchain technology that utilizes sidechains is Aelf~\cite{aelf}.
Aelf takes advantage of multiple chains to improve the efficiency of their 
blockchain, which allows for parallel processing of non-competing transactions.
Sidechains could be the key to achieving scalability, which is essential for 
a blockchain purposed for healthcare.
This is due to the fact that the load that would normally be present on one main 
blockchain could get distributed over the different sidechains, which in turn 
would reduce the possibility of a bottleneck effect occurring on the blockchain network. 

\section{Related Works}\label{section:related}
MedRec \cite{medrec} proposed and deployed a ``decentralized record management 
system to handle EHRs, using blockchain technology''. 
The MedRec open source technology allows patients to achieve ease of access to 
their health records across providers and locations. 
MedRec's permissionless architecture is based off of Ethereum \cite{ethereum}, a well known 
application of blockchain. 
MedRec also touches on using the idea of \textit{data economics} as an 
incentive for mining. 
In contrast, the blockchain discussed in this paper is permissioned and not 
dependent on a cryptocurrency.

MeDShare \cite{medshare} is another blockchain application that 
focuses on the aspect of medical data sharing and the malicious actions that 
could occur with big data entities and data custodians \cite{iothealth}. 
MeDShare's approach primarily focuses on when data is being queried and the 
tracking of all patient data associated with the blockchain. 
MeDShare also implements a way to categorize data sensitivity levels using smart 
contracts. 
The authors claim to have a tamper proof data audit and a way to revoke access 
to data that most blockchain solutions do not have.
Unlike the blockchain in this paper, MeDShare's target users are not patients, 
and user interaction with MeDShare is merely for data requests.

Another proposed framework \cite{dubovit} analyzes aspects of Ethereum and 
Hyperledger \cite{hyperledger} and the different services they offer that 
could be used towards the implementation of a ``secure and trustable'' EHR 
solution using blockchain. 
In \cite{dubovit} the application focuses on cancer patients and the authors' propose a permissioned blockchain that also utilizes smart contracts to control access to data. 
The author’s proposed blockchain uses the Practical Byzantine Fault Tolerance 
(pBFT) consensus protocol which becomes challenging to scale as more nodes 
are added to the network. 
The PoA consensus scheme used in Patient-Healthchain should significantly reduce scalability issues. 

Liu \textit{et al.} \cite{bpds} developed the Blockchain based 
Privacy-Preserving Data Sharing (BPDS) for EHRs. 
BPDS identifies a different way to control patient privacy outside of the 
regular access control mechanisms using a content extraction signature (CES). 
The CES allows patients to censor their medical 
data of identifying information before sharing with other entities. 
Patient-Healthchain does not utilize any access control algorithm at the moment and 
relies on the EHRs being HIPAA compliant with data sharing. 
Although DPoS utilizes less resources to add blocks to the blockchain, the PoA 
consensus scheme discussed in this paper should provide faster block generation.

\section{Relieving the Blockchain Bottleneck for Large Scale Health Care Systems} \label{section:benefits}

\begin{figure*}[t]
    \subfloat[EHR Transactions]{\includegraphics[width=0.49\textwidth, height=3.05in]{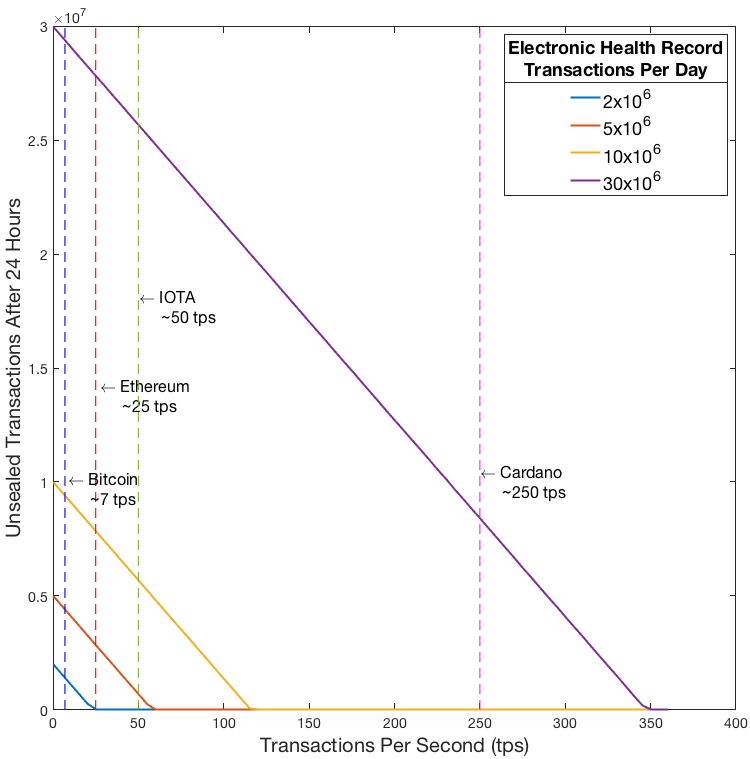}
    \label{fig:graph:ehr}}
    \subfloat[Patient Encounter Transactions]{\includegraphics[width=0.49\textwidth, height=3.05in]{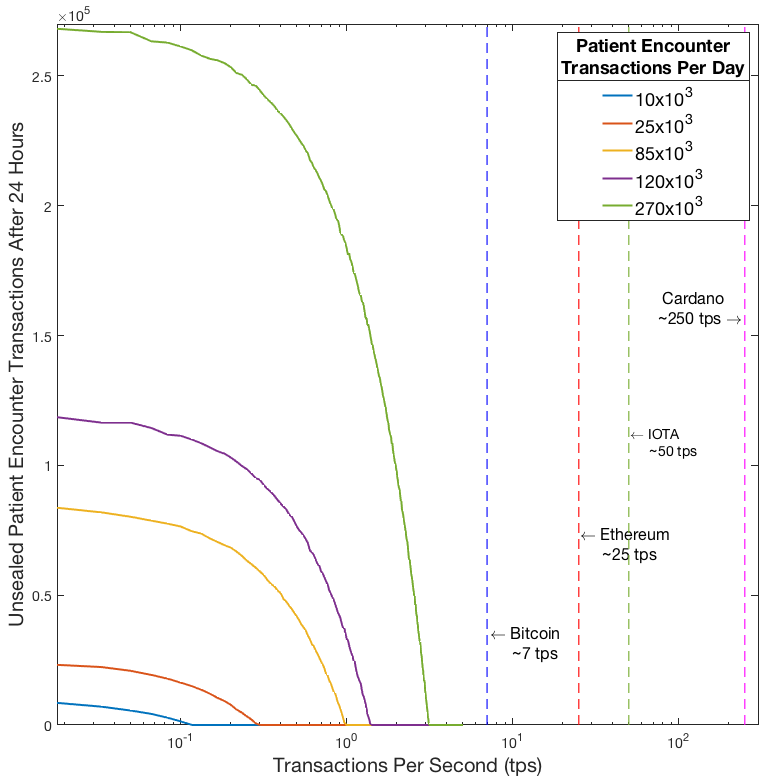}
    \label{fig:graph:patient}}

    \caption{Unsealed Transaction Comparisons}

    \label{fig:graph}
\end{figure*}

While the idea of using blockchain technology in healthcare is not new, there are still 
obstacles that need to be overcome in order for blockchains to be used in such a large 
scale environment.
Approaches have already been proposed in~\cite{medshare, dubovit, bpds} and some have 
implemented such as in~\cite{medrec}.
An important benefit of sidechains is the ability to register transactions and mine 
blocks simultaneously with other sidechains and the parent blockchain. 
The combination of blockchains and sidechains favor healthcare systems because at any 
given time there are multiple patient transactions occurring.
The nature of blockchains requires multiple nodes in the network to come to 
consensus before generating a block, resulting in an eventual bottleneck in 
large scale systems.
Using sidechains that are specific to an individual/patient in the network can impede 
the aforementioned bottleneck on the main blockchain for the following reasons:
\begin{enumerate}
    \item Less transactions would be sent to the mainchain at any given time.
    \item Transactions related to a patient would get added to that respective patient's sidechain, therefore the occurrence of transactions regarding different patients are independent of each other. 
    \item A larger number of transactions could be handled across the network at any given time.
\end{enumerate}

\subsection{Relieving the mainchain of cumbersome transactions}
In Fig.~\ref{fig:graph}, blockchain transactions were simulated to demonstrate 
the current limitations of traditional blockchains and how the utilization of sidechains 
can overcome the problem of bottlenecking when scaling up to the throughput of large 
medical systems. 
In these simulations, Poisson distributions, which models the variation in the number of 
discrete occurrences over a defined interval, were assumed. 
The assumption that a Poisson distribution should be used dates back to the original 
blockchain paper by Satoshi Nakamoto~\cite{nakamoto2008bitcoin} and has been frequently used since~\cite{Rosenfeld2014AnalysisOH}. 
Transactions were generated every second corresponding to the indicated average number of 
transactions per day.
A \textit{sealed} transaction is a transaction that has been added to a successfully hashed block 
on the blockchain. 
Whereas, an \textit{unsealed} transaction is a transaction that is waiting to be added to a block on the 
blockchain.
The abscissa corresponds to the rate per second for sealing transactions into blocks on 
the blockchain. 
The ordinate axis corresponds to the number of unsealed transactions at the end of a 24 
hour period for a given rate of sealing transactions per second. 
Note, the plot of the simulations decrease linearly with the number of sealed 
transactions per second. 
Where the plot reaches the x-axis indicates the rate of sealing transactions necessary 
to handle the throughput for a given average rate of generating transactions.

Fig.~\ref{fig:graph:ehr} shows the results if every computerized transaction for a 
electronic health record system were sealed in blocks on the main blockchain. 
\footnote{S. O'Neal, ``Who Scales It Best? Inside Blockchains' Ongoing Transactions-Per-Second Race,'' Cointelegraph, 2019, \url{https://cointelegraph.com/news/who-scales-it-best-inside-blockchains-ongoing-transactions-per-second-race}, Accessed: 10/3/19}
This is equivalent to traditional blockchains attempting to handle all transactions.
According to University of Kentucky (UK) HealthCare,\footnote{L. Dawahare, ``UK HealthCare Exploring Ways Big Data Analytics Can Improve 
Patient Care,'' UK College of Medicine, 2015,
\url{https://med.uky.edu/news/uk-healthcare-exploring-ways-big-data-analytics-can-improve-patient-care}, Accessed: 10/3/19} the average number of computerized transactions for a large medical system is 
approximately 10 million per day. 
Neither Bitcoin, Ethereum, or IOTA can handle the throughput of a system this large as they 
currently support $~7$ tps, $~25$ tps, and $~50$ tps respectively. 
When looking at Ethereum in Fig.~\ref{fig:graph:ehr}, it can be seen that when attempting 
to seal the 10M transactions of the UK HealthCare system, over 7.5 transactions are left unsealed at the end of each day.
The largest number of EHR transactions per day that this study found in a 
real-world setting was 30 million per day\footnote{N. Morris, ``Change Healthcare: enterprise blockchain with 30 million transactions per day,'' Ledger Insights, 2019, \url{https://www.ledgerinsights.com/change-healthcare-enterprise-blockchain-with-30-million-transactions-per-day/}, Accessed: 10/3/19}, for which Bitcoin, Ethereum, IOTA, and Cardano leave a substantial number of transactions 
unsealed at the end of a 24 hour period. 
Ethereum, becomes a viable option only after the 
throughput drops below approximately 2 million transactions per day. 
One might think to increase the block sizes or create more frequent block events to increase 
transaction throughput, but this results in more conflicts between blocks and reduces the 
level of security from attacks~\cite{sompolinsky2015secure}. 

Fig.~\ref{fig:graph:patient} shows the results if individual patient encounters were 
used as transactions on the main blockchain. A patient encounter transaction includes any in-patient discharge summary, any out-patient doctor visit or surgery. The x-axis is plotted on a logarithmic sale because the range of the number of patient encounters per second in real healthcare systems span several orders of magnitude.
Kaiser Permanente, one of the largest healthcare systems, 2018 Annual Report shows they 
have 47M doctor visits a year~\cite{kaiserData}.
Another report says Kaiser Permanente has over 100M patient encounters with physicians 
a year\footnote{D. Barkholz, ``Kaiser Permanente chief says members are flocking to virtual visits,'' Modern Healthcare, 2017, \url{https://www.modernhealthcare.com/article/20170421/NEWS/170429950/kaiser-permanente-chief-says-members-are-flocking-to-virtual-visits}, Accessed: 10/3/19}.
40M patient encounters and 100M doctor visits translates to approximately 128,767 and 273,973 patient encounters per day, respectively. %
The slowest of the blockchains; Bitcoin and Ethereum can handle the throughput of several healthcare systems the 
size of Kaiser if only encounters are maintained on the main blockchain.
Here, we assume transactions related to inputting data to individual patient records can
be handled by sidechains which is what is proposed in Section~\ref{section:architecture}.

\begin{table}%
    \centering
    \caption{Number of Sidechains Needed}
    \resizebox{0.48\textwidth}{!}{
    \begin{tabular}{|c|c|c|c|c|c|} %
        \hline
        \multirow{2}{*}{\textbf{Number of Patients}} &  
        \multirow{2}{*}{\textbf{Number of Transactions Per Day}} &
        \multicolumn{4}{c|}{\textbf{Number of Sidechains Needed}} \\\cline{3-6} 
        & & \multicolumn{1}{|c|}{\textbf{Bitcoin}} & \multicolumn{1}{c|}{\textbf{Ethereum}} &
        \multicolumn{1}{c|}{\textbf{IOTA}} &
        \multicolumn{1}{c|}{\textbf{Cardano}} \\
        \hline
        1 & 110 &  1 & 1 & 1 & 1 \\
        \hline 
        1,000 & 110,000 & 1 & 1 & 1 & 1 \\
        \hline
        10,000 & 1,100,000 & 2 & 1 & 1 & 1 \\
        \hline
        50,000 & 5,500,000 & 10 & 3 & 2 & 1 \\
        \hline
        100,000 & 11,000,000 & 19 & 6 & 3 & 1 \\
        \hline
        200,000 & 22,000,000 & 37 & 11 & 6 & 1 \\
        \hline
        300,000 & 33,000,000 & 55 & 16 & 8 & 2 \\
        \hline
    \end{tabular}}
    \label{table:results_summary}
\end{table}

\subsection{Scaling with sidechains}
A natural succeeding question is, ``how many sidechains should be generated to support 
mainchains for EHR systems?''
The amount of sidechains needed is not easy to answer as the type of sidechain and 
respective consensus mechanism selected significantly impact the resources needed to 
operate. 
Table~\ref{table:results_summary} provides insight into how sidechains might handle the 
bandwidth of transactions depending on how many patients are placed on a sidechain. 
We do not propose a particular type of blockchain be used for sidechains, but instead 
show the amount of patients some traditional blockchains can support. 

To determine the average amount of transactions per patient per day, we use the 30M 
transactions mentioned earlier and divide it by the total number of patient encounters at 
Kaisier Permanente per day, 273,973, to get $\sim$110 transactions. 
Note that this is most likely an overestimate for many hospital systems as we used the 
largest amount of transactions found and patient encounters for a large hospital system 
such as Kaiser. 
If the amount of transactions per patient per day for a particular hospital system ends 
up being less than 110, a lower amount of sidechains will be needed than displayed 
in Table~\ref{table:results_summary}.
We also assume that each of the 100M patient encounters reported by Kaisier Permanente 
per year is a unique patient, which may not necessarily be true.

\section{Patient-Healthchain Architecture} \label{section:architecture}
This section of the paper covers the Patient-Healthchain architecture that uses sidechains for Healthcare systems. 
Patient-Healthchain allows patients to have one central place to access all their medical 
information regardless of the underlying system different hospitals use. 
Having a system that appears uniform to the patient should help improve the overall 
perception about technology for healthcare.

\subsection{Blockchain Members and Transactions} \label{subsection:members}
Patient-Healthchain is a private permissioned blockchain, also known as a consortium 
blockchain. 
In a permissioned blockchain only members of the 
consortium (e.g. hospitals) and users that consortium member permits (e.g. patients) will be able to join the blockchain. 
For the scope of this paper, blockchain members are hospitals and users are patients, future 
work might determine that it is beneficial to allow researchers or insurance adjusters as members or users.

\begin{table}%
    \centering
    \caption{Transactions}
    \begin{tabular}{|p{4.0cm}|p{1.8cm}|p{1.2cm}|} %
        \hline
        \multicolumn{1}{|c|}{\textbf{Transaction Types}} &  \multicolumn{1}{c|}{\textbf{Members}} &  \multicolumn{1}{c|}{\textbf{Chain}}\\
        \hline
        Joining/Leaving the network. & Hospital/Patient & Mainchain \\
        \hline 
        End of patient interactions/visit that produce discharge summaries & Hospital/Patient & Mainchain, Sidechain \\
        \hline
        Information sharing with other hospital systems in the consortium & Hospital & Sidechain \\
        \hline
        Health records being accessed by a  new entity & Patient & Sidechain \\ 
        \hline
        New diagnoses from medical staff/Changes in health records & Hospital/Patient & Sidechain \\
        \hline
        Financial requests/Transactions & Hospital/Patient & Sidechain \\
        \hline
    \end{tabular}%
    \label{tab:transactions}
\end{table}

In Patient-Healthchain, a transaction can be the record of any interaction between a 
patient and their medical provider, an interaction between hospital systems, or a change 
to a patient's health record. 
When a member joins or leaves Patient-Healthchain, it is also recorded as a transaction. 
Depending on the type of transaction, a transaction will either get published to the main 
blockchain or to a hospital/patient sidechain. 
Specific transactions are outlined in Table~\ref{tab:transactions}.

\subsection{Mainchain-Sidechain Relationship and Block Structure} \label{subsection:components}%
\begin{figure}
  \begin{center}
    \includegraphics[width=\columnwidth, height=2.5in]{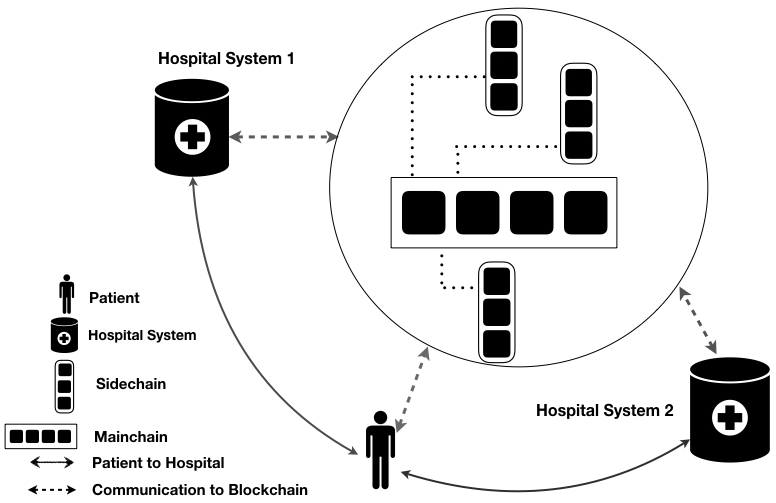}
  \end{center}
  \caption{Architecture Overview}
  \label{fig:arch}
 \end{figure}

Fig.~\ref{fig:arch} depicts the Patient-Healthchain Architecture and the relationships between entities in the network. 
Patient-Healthchain assumes that patients interact with hospitals that do not naturally
communicate with each other.
Individual hospital systems can, however, communicate with the blockchain network, 
enabling them to interact with each other. 
Patients will also be able to interact with hospitals through the blockchain network.
The structure of Patient-Healthchain consists of a ``mainchain'' and multiple 
``sidechains''. 
A sidechain is assigned to its respective member when the member initially joins Patient-Healthchain. %
When a patient joins the blockchain network a link is created between the patient's 
initial block on the mainchain and their newly created sidechain; the same process 
occurs when a hospital joins the network. 
The purpose of a patient member's sidechain is to retain all information relevant to a specific patient's medical lifecycle. 
A hospital's sidechain functions as a standalone blockchain for the hospital system. %
 
\begin{figure}
  \begin{center}
    \includegraphics[width=\columnwidth,height=1.5in]{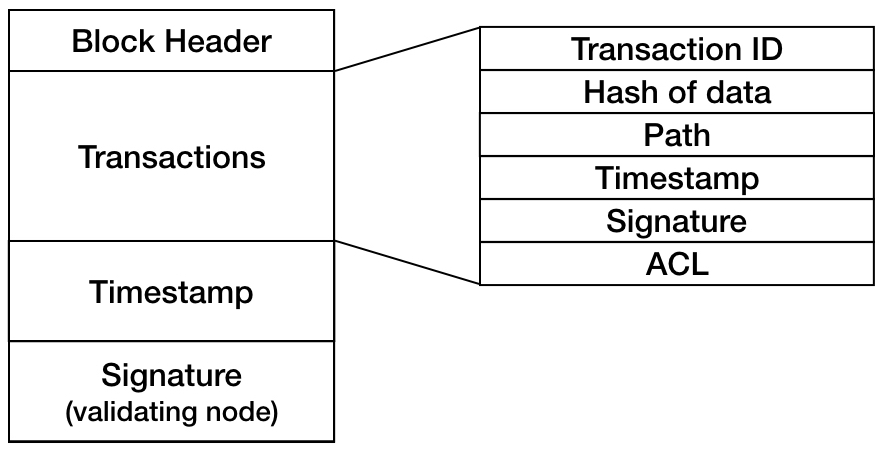}
  \end{center}
  \caption{Block Overview}
  \label{fig:block}
 \end{figure}

Fig.~\ref{fig:block} shows the types of data that makes up a block in 
Patient-Healthchain. 
A block could be composed of multiple transactions. %
The \textit{block header} is the hash of the header of the preceding block. 
\textit{Transactions} consist of recorded information between members and have the 
following attributes: 
\begin{itemize}
    \item Transaction ID --- identifies the transaction and link same transactions between mainchain and sidechains.
    \item Hash of data --- used to verify that the data being accessed (usually patient medical information) has not been tampered with.
    \item Path --- path to related data (access purposes) 
    \item Timestamp --- time that the related transaction occurred.
    \item Signatures --- signatures from related parties to verify that data is accurate. (e.g. patient, doctor)
    \item Access Control List (ACL) --- managed by smart contracts, will contain a list of permitted public keys that are authorized to access information related to the respective transaction.
\end{itemize}
The block \textit{timestamp} will refer to the time that the block was created. 
The block \textit{signature} is the signature of the validating node who published that 
block. 
Block structures remain consistent on both the mainchain and the sidechains. 

\subsection{Proof of Authority for Patient-Healthchain} \label{subsection:PoA}
As discussed in Section \ref{subsection:consensus}, in order to choose a fitting 
consensus scheme for a blockchain used in healthcare, the following should hold: the 
consensus scheme should not consume a lot of power, throughput to the blockchain should 
be able to keep up with the average throughput of a healthcare system, resulting in a blockchain that 
scales to large system use.
A consensus scheme that consists of the aforementioned characteristics is PoA.
The PoA consensus scheme can work well for blockchains in healthcare systems for a 
number of reasons, including:
\begin{enumerate}
    \item  Validator's identities can be checked and confirmed by public notaries. 
    Confirming an identity ensures that a member is who they say they are and that the 
    member will be completely liable for actions that could jeopardize the privacy and 
    security of the blockchain.
    \item The blockchain is private, therefore it is only possible for members of the 
    consortium to be able to become a validator and generate blocks.
    \item  Due to the nature of the data that is being shared over the blockchain, 
    patient trust and the trust between other consortium members would increase knowing 
    that any bad actor in the blockchain network can be easily identified and 
    reprimanded. 
    \item The throughput of a blockchain for healthcare should be equal to or faster 
    than what a patient or provider expects/needs in comparison to the current system 
    in place. 
    PoA removes the computation component of other consensus mechanisms and allows for 
    almost immediate generation of blocks to the chain.
    \item PoA also allows for the absence of cryptocurrency on the blockchain. 
    The incentive for use is the overall ease of patient data sharing across multiple 
    facilities, e.g. increased interoperability, which has proven to be an issue for 
    providers \cite{emrChallenges}. %
\end{enumerate}

\subsection{Patient Consideration for EHRs}%
Patient-Healthchain aims to put access control of records in the hands of patients as well
as help integrate different EHRs at the provider level.
EHRs should also always be up-to-date and readily available to the patients.
There has been advancement in this area, for example early in 2018 Apple announced that
12 hospitals were participating in merging their EHRs with Apple’s Health Records 
app\footnote{Apple Inc., ``Apple announces effortless solution bringing health records to iPhone,'' Apple Newsroom, 2018, \url{https://www.apple.com/newsroom/2018/01/apple-announces-effortless-solution-bringing-health-records-to-iPhone/}, Accessed: 10/3/19}. %
Patients are able to synchronize EHR data from multiple healthcare systems to their phone 
for ready access.
However, the data synchronized into the Health Records app is not available for 
providers, accordingly if a patient wants to share information from one health care 
system to another they must manually show their data to providers\footnote{L. Mearian, ``A tale of 
two hospitals that adopted Apple’s Health Record app,'' Computer World, 2018, \url{https://www.computerworld.com/article/3268467/a-tale-of-two-hospitals-that-adopted-apples-health-record-app.html}, Accessed: 10/3/19}.
Facilitating the transfer of information between providers using different hospital 
systems can be handled using the proposed Patient-Healthchain architecture.
A patient's respective sidechain provides them with the ability to see their whole EHR 
along with controlling the providers who have access to their information.
Access control is handled through an authorized list of provider keys that are updated 
via smart contracts whenever a patient grants or revokes access.
Patients are able to see when providers are requesting and sharing information, and if 
their provider changes, the patient can also choose to revoke access to their data.

\section{Conclusion} \label{section:conclusion}
This paper quantifies and puts into perspective the bottleneck problem that could occur 
when trying to use a single blockchain to handle multiple healthcare systems and their 
patients due to the inability for current blockchain architectures to keep up with the 
output of transactions in healthcare.
We also propose relieving bottlenecks on the blockchain by using sidechains.
Lastly, the Patient-Healthchain architecture is proposed as an advancement towards solving 
the EHR interoperability problem and the issues surrounding a bottleneck. 
Future work will explore the types of workload being placed on the sidechains,
particularly when it comes to different consensus schemes and the burden placed on miners that results in energy consumption.

\bibliographystyle{IEEEtran}
\bibliography{sample-base}

\end{document}